\documentclass[twocolumn,preprintnumbers,superscriptaddress,landscape,nofootinbib]{revtex4}
\usepackage[utf8]{inputenc}
\usepackage{amsmath}
\usepackage{amsthm}
\usepackage{amssymb}
\usepackage{float}
\usepackage{braket}
\usepackage{graphicx}
\usepackage{url}
\usepackage{here}
\usepackage{natbib}
\usepackage{rotating}
\usepackage[colorlinks=true, pdfstartview=FitV, linkcolor=red, citecolor=blue, urlcolor=blue]{hyperref}
\usepackage{slashed}
\usepackage{multirow}
\usepackage[normalem]{ulem}
\graphicspath{{./figures/}}

\newcommand{\be}{\begin{equation}}      
\newcommand{\ee}{\end{equation}}      
\newcommand{\bea}{\begin{eqnarray}}      
\newcommand{\eea}{\end{eqnarray}}

\newcommand{\Tr}{\mathrm{Tr}}

\newcommand{\ctext}[1]{\raise0.2ex\hbox{\textcircled{\scriptsize{#1}}}}

\usepackage{tikz}
\usetikzlibrary{quantikz}

\theoremstyle{definition}

\theoremstyle{remark}

\usepackage[normalem]{ulem}

\begin{document}

\title{Long-range quantum energy teleportation and distribution on a hyperbolic quantum network} 
\author{Kazuki Ikeda}
\email[]{kazuki7131@gmail.com}
\affiliation{Co-design Center for Quantum Advantage, Stony Brook University, Stony Brook, New York 11794-3800, USA}
\affiliation{Center for Nuclear Theory, Department of Physics and Astronomy, Stony Brook University, Stony Brook, New York 11794-3800, USA}

\bibliographystyle{unsrt}

\begin{abstract}
Teleporting energy to remote locations is new challenge for quantum information science and technology. Developing a method for transferring local energy in laboratory systems to remote locations will enable non-trivial energy flows in quantum networks. From the perspective of quantum information engineering, we propose a method for distributing local energy to a large number of remote nodes using hyperbolic geometry. Hyperbolic networks are suitable for energy allocation in large quantum networks since the number of nodes grows exponentially. To realise long-range quantum energy teleportation, we propose a hybrid method of quantum state telepotation and quantum energy teleportation. By transmitting local quantum information through quantum teleportation and performing conditional operations on that information, quantum energy teleportation can theoretically be realized independent of geographical distance. The method we present will provide new insights into new applications of future large-scale quantum networks and potential applications of quantum physics to information engineering.
\end{abstract}

\maketitle

\section{Hyperbolic Geometry for Quantum Energy Distribution}
Optimal network design is crucial to the efficient distribution of information. Hyperbolic geometric networks are practical and beneficial as a design for distributing resources to many nodes~\cite{Wang_Zhang_Shi_2019,PhysRevE.82.036106}. The number of nodes in a Euclidean network increases only on the order of a polynomial with distance, but that of nodes in a hyperbolic network increases exponentially with distance from the center. Given that hyperbolic network design has been very successful in classical information processing, adapting this idea to quantum information processing is the right direction. 

There is a hyperbolic-design quantum processor based on cQED~\cite{kollar2019hyperbolic}, which motivates various active studies in condensed matter physics and meta-material science~\cite{2020arXiv200805489M,doi:10.1073/pnas.2116869119,2021arXiv210710586I,Ikeda_2021,PhysRevE.106.034114,PhysRevB.105.125118}. In this study, we propose a method of quantum energy teleportation (QET) and distribution (QED) in hyperbolic quantum networks. The method of this study presents a new way to utilize quantum information processing of such hyperbolic geometrically designed processors. QET is a method of transferring energy to a remote location using only LOCCs, with entangled ground states in a quantum many-body system.

While teleportation of quantum states on the quantum has already been established theoretically and experimentally~\cite{PhysRevLett.70.1895,furusawa1998unconditional,2015NaPho...9..641P,takeda2013deterministic}, the possibility of teleportation with respect to energy is relatively unexplored~\cite{HOTTA20085671,2009JPSJ...78c4001H,2015JPhA...48q5302T,2009PhRvA..80d2323H,PhysRevA.84.032336,PhysRevA.82.042329,Hotta_2010,2023arXiv230102666I}. The first objective of this study is to establish a method for transferring the energy of a quantum many-body system in a laboratory to a spatially distant location. This idea can be achieved quite easily by combining it with conventional quantum state teleportation (QST). First of all, energy is an eigenvalue of the Hamiltonian, but only initial states, unary operations, and measurements constitute a quantum circuit, and even the concept of a Hamiltonian is not given a priori to execute a quantum algorithm on quantum computers/processors. In particular, the Hamiltonian is involved in QET only in the ground state, which is prepared as the initial state of the QET protocol. Once the ground state is prepared, all that remains is to follow the protocol and observe the state after local operations and classical communication to obtain a nontrivial energy flow. 

In QET, information about a given Hamiltonain is needed for conditional operations (especially the rotation angle) and ground state design. Designing the correct ground state is indeed a difficult problem, and it is sometimes extremely difficult (as hard as QMA) to find the ground state of the general Ising model. However, unlike solving general physics problems, it is not difficult for QET to construct a good ground state. In the simplest case, only 2 qubit states are needed. As we present in this work, there is no need to use a huge number of qubits even for large-scale QET and QED, using the property of hyperbolic graphs that the number of nodes grows exponentially. In other words, even if the ground state cannot be obtained analytically, one can easily implement a quantum circuit that reproduces it by some numerical methods (this is possible by IBM Qiskit package, for example). Thus, QET and QED are very interesting as a problem of pure information engineering, although their foundation is physics.

\begin{figure*}
    \centering
    \includegraphics[width=\linewidth]{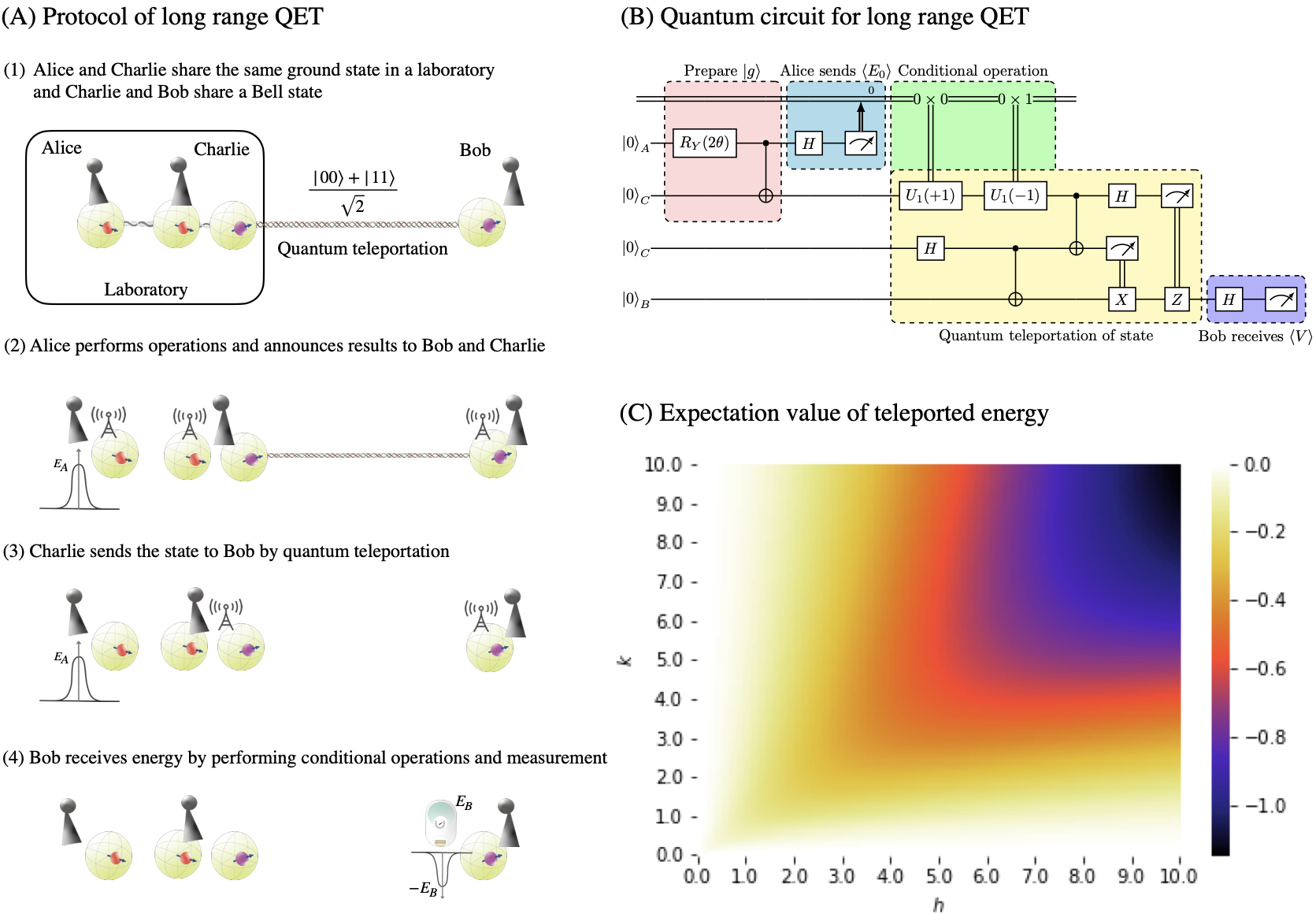}
    \caption{Complete protocol (A) and quantum circuit (B) for long-range quantum energy teleportation. Alice announces her measurement result to Bob and Charlie. Bob delegates the conditional operation $U_{1}(\mu)$ to Charlie who is spatially very close to Alice and teleports quantum states to Bob. Then Bob can receive $\langle V\rangle$ statistically by measuring his local qubit in $X$ basis. (C) Parameter dependence of theoretical expectation value of teleported energy $\langle E_B\rangle=\Tr[\rho_\text{QET}H_{n_B}]$ to Bob's local system. Bob will receive $-\langle E_B\rangle$ through his measurement device.}
    \label{fig:complete}
\end{figure*}
\section{\label{sec:long}Long Distance QET by Quantum State Teleportation}
The purpose of this section is to generalize the quantum circuit given in previous work by the author~\cite{2023arXiv230102666I} to a long-range QET. The complete quantum circuit is given in Fig.~\ref{fig:complete}. Let us consider Alice sending energy to Bob, who is far away. Bob entrusts the conditional operation to Charlie, who is very close to Alice spatially, and Alice tells her measurement result to both Charlie and Bob first. Bob and Charlie share a Bell state $\frac{\ket{00}+\ket{11}}{\sqrt{2}}$, and Charlie uses the entanglement to transfer the quantum state to Bob. The part framed in yellow in Fig.~\ref{fig:complete} corresponds to the quantum circuit that performs quantum state teleportation. At this time, Charlie only performs standard QST and does not gain energy. Bob can obtain the energy sent by Alice by performing the necessary operation and measurement on the state transferred from Charlie. For example, in Fig.~\ref{fig:complete}, Bob can statistically obtain $\langle V\rangle$ by observing his own $X$ operator.

Let $k,h$ be positive real numbers. The Hamiltonian of the minimal model is
\begin{align}
    H_\text{tot}&=H_0+H_1+V,\\
    H_n&=hZ_n+\frac{h^2}{\sqrt{h^2+k^2}},~(n=0,1)\\
    V&=2kX_0X_1+\frac{2k^2}{\sqrt{h^2+k^2}}.
\end{align}
The ground state of $H_\text{tot}$ is 
\begin{equation}
\label{eq:groundstate}
    \ket{g}=\frac{1}{\sqrt{2}}\sqrt{1-\frac{h}{\sqrt{h^2+k^2}}}\ket{00}-\frac{1}{\sqrt{2}}\sqrt{1+\frac{h}{\sqrt{h^2+k^2}}}\ket{11},
\end{equation}

The constant terms in the Hamiltonians are added so that the ground state $\ket{g}$ of $H_\text{tot}$ returns the zero mean energy for all local and global Hamiltonians: 
\begin{equation}
\label{eq:ground}
\bra{g}H_\text{tot}\ket{g}=\bra{g}H_0\ket{g}=\bra{g}H_1\ket{g}=\bra{g}V\ket{g}=0. 
\end{equation}
However, it should be noted that $\ket{g}$ is neither a ground state nor an eigenstate of $H_n,V, H_n+V~(n=0,1)$. The essence of QET is to extract negative ground state energy of those local and semi-local Hamiltonians. 

The long-range QET protocol is as follows. We consider the situation where Alice sends energy to Bob at a distance via Charlie, who is close to Alice. First, Alice measures her Pauli operator $X_0$ by $P_0(\mu)=\frac{1}{2}(1+\mu X_0)$ and then she obtains either $\mu=-1$ or $+1$. It turns out that Alice's expectation energy is $E_0=\frac{h^2}{\sqrt{h^2+k^2}}$.

Via a classical channel, Alice then sends her measurement result $\mu$ to Bob and Charlie. Then Charlie applies an operation $U_1(\mu)$ to his qubit and measures $H_1$ and $V$. Then Charlie and Bob share a Bell state $\frac{\ket{00}+\ket{11}}{\sqrt{2}}$ and Charlie sends his state $U_1\ket{A_\mu}$ to Bob by QST. As is well known, Charlie can send any quantum state to Bob without knowledge of the quantum state he possesses.
The density matrix $\rho_\text{QET}$ that Bob receives after Charlie operates $U_1(\mu)$ to $P_0(\mu)\ket{g}$ is 
\begin{equation}
\label{eq:rho_QET}
    \rho_\text{QET}=\sum_{\mu\in\{-1,1\}}U_1(\mu)P_0(\mu)\ket{g}\bra{g}P_0(\mu)U^\dagger_1(\mu). 
\end{equation}
Using $\rho_\text{QET}$, the expected local energy at Bob's subsystem is evaluated as $\langle E_1\rangle=\Tr[\rho_\text{QET}(H_1+V)]$, which is negative in general. Due to the conservation of energy, $ E_B=-\langle E_1\rangle (>0)$ is extracted from the system by the device that operates $U_1(\mu)$~\cite{PhysRevD.78.045006}. In this way, Alice and Bob can transfer the energy of a quantum system in a laboratory by LOCC only, without geographical constraints.

It is certainly (easily) possible to realize long-distance QET, by combing QST. Preparing the initial ground state $\ket{g}$ is easily possible with a 2-qubit processor, and the technology to transfer the state far enough away by QST was established more than 25 years ago~\cite{furusawa1998unconditional,2015NaPho...9..641P,takeda2013deterministic}. Even if the initial state is more than two qubits, quantum information processing techniques including quantum teleportation can be combined to remotely transfer quantum states generated locally in quantum devices/hardware/materials. Since QET can be realized by simply performing LOCC, the discussion below proceeds under the assumption that long-range QET is unconditionally possible. In other words, QET is difficult to perform only when it is difficult to generate a suitable initial state and when it is difficult to perform quantum teleportation on a quantum network. From this perspective, QET is an interesting new challenge for information engineering and quantum information processing.

\begin{figure*}
    \centering
    \includegraphics[width=\linewidth]{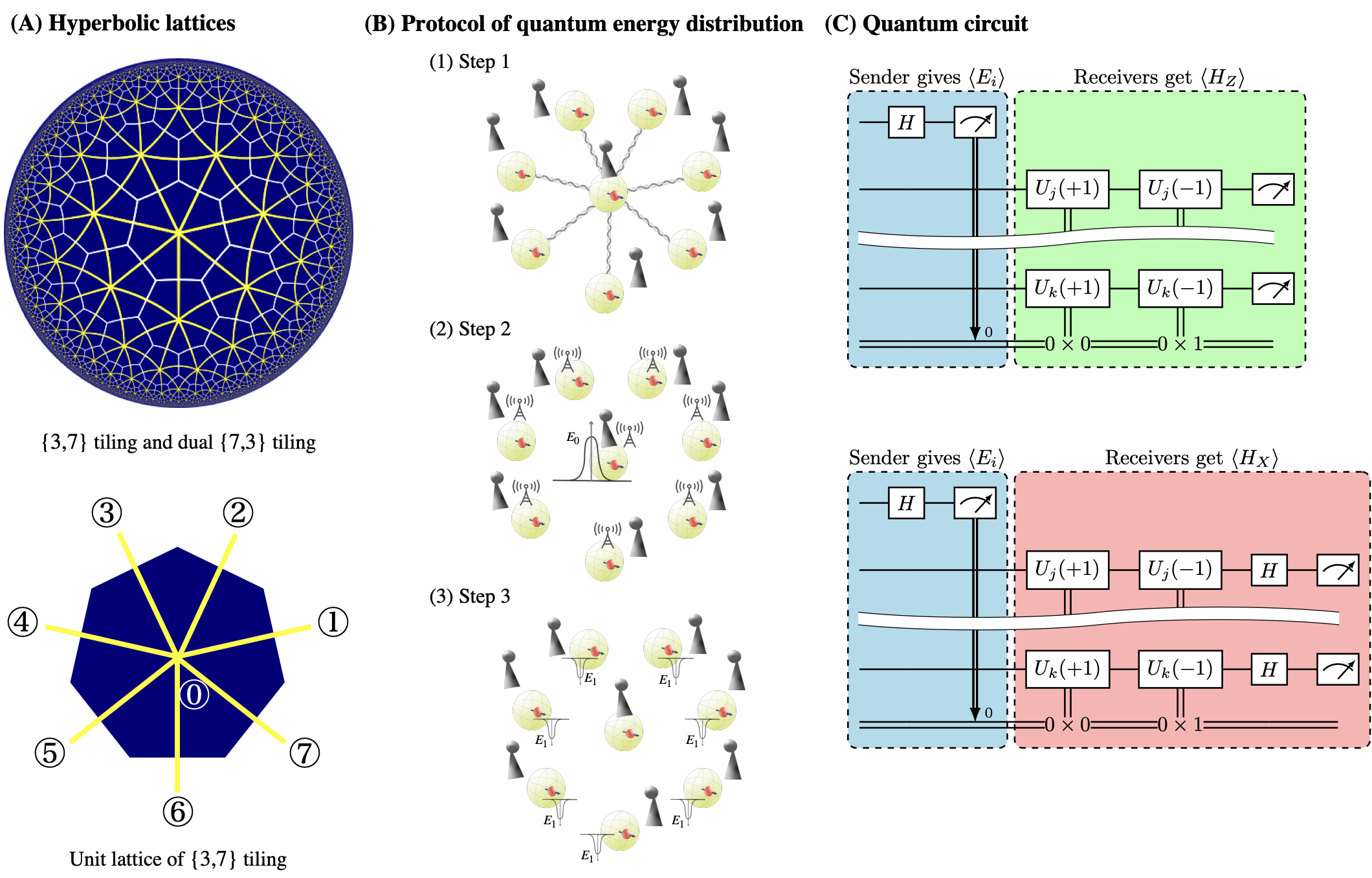}
    \caption{(A) [Upper] Hyperbolic lattice [Lower] unit lattice (B) Protocol of quantum energy distribution;  Alice measures $X_0$ and tells her result  $\mu\in\{\pm1\}$ to people on the network via classical communication. They can receive energy by applying conditional operation $U_1(\mu)$ and measuring $Z_j$ and $X_j$. (C) Quantum circuits to implement the protocol given in (B).}
    \label{fig:operation}
\end{figure*}
\begin{figure*}
    \centering
    \includegraphics[width=\linewidth]{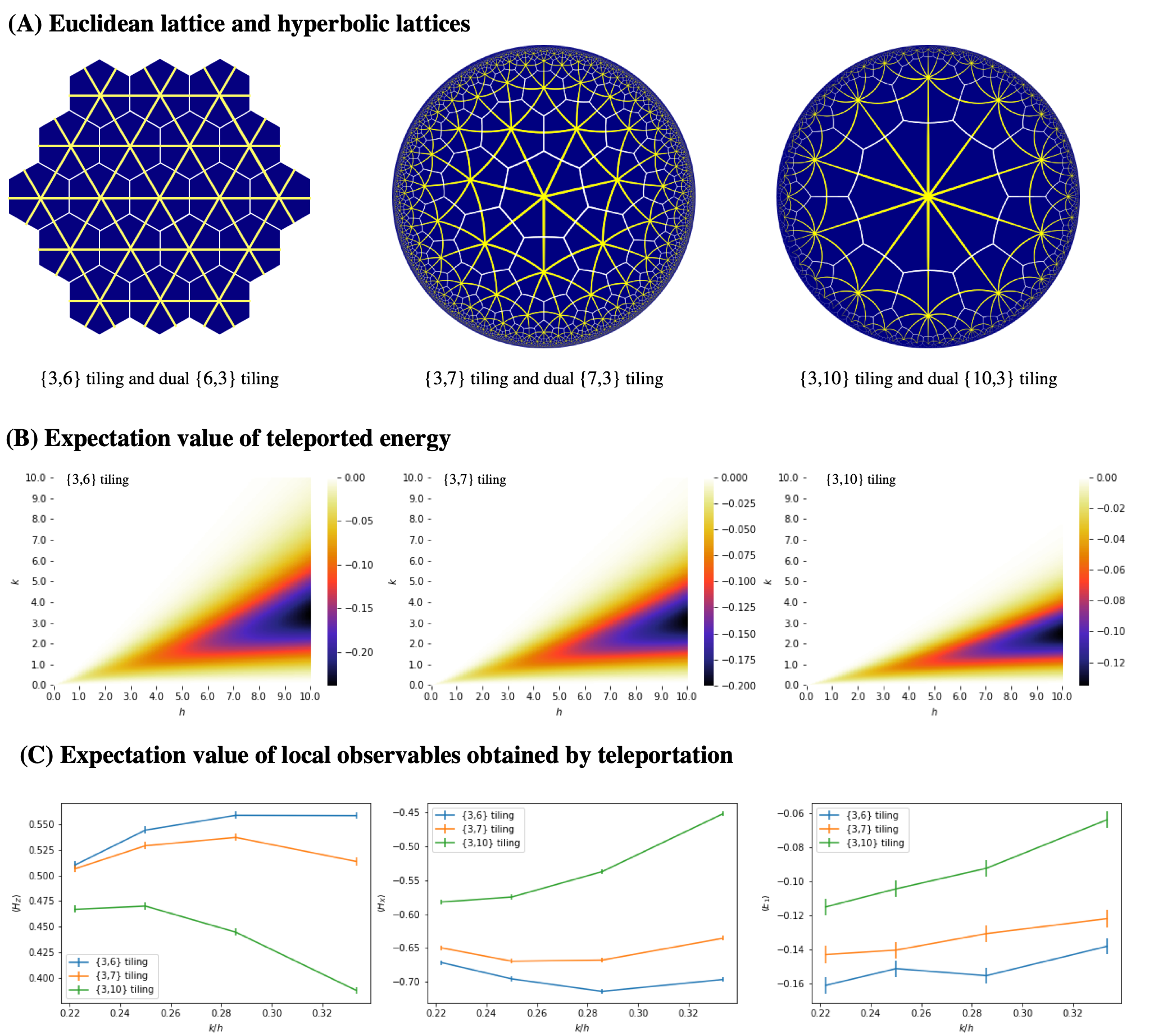}
    \caption{(A) [Left] Euclidian lattice of $\{3,6\}$ tiling (yellow) and dual lattice of $\{6,3\}$ tiling (white). [Middle] Hyperbolic lattice of $\{3,7\}$ tiling (yellow) and dual lattice of $\{7,3\}$ tiling (white). [Right] Hyperbolic lattice of $\{3,10\}$ tiling (yellow) and dual lattice of $\{10,3\}$ tiling (white). (B) Expected energy teleported to a node in the unit lattice of each tiling of $\{3,6\},\{3,7\},\{3,10\}$. (C) Simulation of measurement of local operators $X,Z$, evaluated by $10^5$ samplings. Error bars correspond to statistical errors.}
    \label{fig:my_label}
\end{figure*}
\section{\label{sec:Long}Design of quantum energy distribution on hyperbolic quantum networks}
Here let us consider QET/QED on a network consisting of an exponentially large number of participants in a network. This situation is quite natural given the circumstance of our (classical) networked society today. In this network, it is of utmost importance to deliver quantum resources instantaneously to the vast number of nodes that make up the network. For this, let $L=(V,E,F)$ be a hyperbolic lattice determined by a set of vertices $V$, a set of edges $E$ and a set of lattice faces $F$.
The hyperbolic lattices we use are specified with $\{3,q\}$ tiling; they are tessellations of the plane consisting of regular triangles such that each vertex is connected to $q$ vertices. In general, one can consider $\{p,q\}$ tiling, which are hyperbolic when $(p-2)(q-2)>4$, otherwise Euclidean. Therefore $\{3,7\}$ tiling is the simplest hyperbolic lattice we consider in this work. For $\{3,q\}$ tiling ($q\ge 6$), the Hamiltonian is 
\begin{align}
    H_{Z,i}&=hZ_i+\epsilon,~(i=0,\cdots,q)\\
    H_{X,j}&=kX_0X_j+\varepsilon,~(j=1,\cdots,q)\\
    H_\text{hyp}&=\sum_{i=0}^qH_{Z,i}+\sum_{j=1}^qH_{X,j}
\end{align}
A tiling $\{3,q\}$ is hyperbolic when $q\ge 7$ and it is Euclidean when $q=6$, as shown in Fig.~\ref{fig:my_label}. Each $\epsilon_i$ should be chosen in such a way that 
\begin{align}
\begin{aligned}
    \bra{g}H_\text{hyp}\ket{g}=\bra{g}H_i\ket{g}=0,~\forall i\in E
\end{aligned}    
\end{align}
where $\ket{g}$ is the ground state of the total Hamiltonian $H_\text{hyp}$. In general, $\ket{g}$ is not the ground state of local $H_i$. 

We assign a participant on each vertex of $L$ and they send/receive energy. A sender at $i=0$ of energy performs a projective measurement with $P_0(\mu)=\frac{1}{2}(1+\mu\sigma_0)$ and announces $\mu\in\{-1,+1\}$ to people in $N_0$. Then a receiver at $j$ can receive energy by applying conditional operation $U_j(\mu)$ and measuring $Z_j$ and $X_j$.

In what follows let us describe more on those operators of a sender and a receiver.  We define $U_j(\mu)$ by 
\begin{equation}
    U_{j}(\mu)=\cos\theta I-i\mu\sin\theta\sigma_j,
\end{equation}
where $\theta$ obeys 
\begin{align}
    \cos(2\theta)&=\frac{\xi}{\sqrt{\xi^2+\eta^2}}\\
    \sin(2\theta)&=-\frac{\eta}{\sqrt{\xi^2+\eta^2}}
\end{align}
where $\xi$ and $\eta$ are 
\begin{align}
    \xi&=\bra{g}\sigma_jH\sigma_j\ket{g}\\
    \eta&=\bra{g}\sigma_i\dot{\sigma}_j\ket{g}
\end{align}
with $\dot{\sigma_j}=i[H_{j},\sigma_j]=i[H,\sigma_j]$. The average quantum state $\rho_\text{QET}$ is obtained after Bob operates $U_{j}(\mu)$ to $P_{i}(\mu)\ket{g}$. Then the average energy a receiver gains is 
\begin{equation}
\label{eq:QET}
    \langle E_{j}\rangle=\Tr[\rho_\text{QET}H_{j}],
\end{equation}
where $\rho_\text{QET}$ is 
\begin{equation}
\label{eq:rho_QET}
    \rho_\text{QET}=\sum_{\mu\in\{-1,1\}}U_j(\mu)P_0(\mu)\ket{g}\bra{g}P_0(\mu)U^\dagger_j(\mu). 
\end{equation}
Note that the operation of one receiver does not affect the results of another receiver since $[U_j,U_k]=0$. In what follows we use $\sigma_i=X_0$ for the sender's operation and $\sigma_j=Y_j~(j=1,\cdots,q)$ for the receiver's operation. 

Suppose a sender (Alice) at $i=0$ wants to distribute her energy to a receiver (Bob) at distance of more than 1. In this case, by inviting a third party (Charlie), Alice and sends her energy to Bob without creating a new direct interaction between Alice and Bob. We assume that Charlie is faithful and trustworthy enough. Even in this situation, as described in Sec.~\ref{sec:long}, Bob can outsource his local operations to Charlie and have him send the quantum state by QST. Then Bob can obtain the energy by performing the necessary operations and measurements on the sent state. Hence, as we have repeatedly stated, QET and QED can be performed on hyperbolic lattices without geographic constraints by combining QST. Since QST can also be realized using only LOCC, all protocols are completed using only LOCC.

In Fig.~\ref{fig:my_label} and Table~\ref{tab:processor}, we present our results of QET and QED on both a Euclidian network and hyperbolic networks. Fig.~\ref{fig:my_label} (A) shows the geometry of the networks we consider. The hyperbolic lattices we used are specified with the $\{3,7\}$ tiling and the $\{3,10\}$ tiling. For comparison, we use the honeycomb lattice specified with the $\{3,6\}$ tiling. The per capita energy that a participant on a unit lattice can receive when using each lattice is shown in Fig.~\ref{fig:my_label} (B). The energy expectation value of each operator is shown in  Fig.~\ref{fig:my_label} (C). To demonstrate QET and QED on both Euclidean and hyperbolic lattices, we performed quantum simulation using qasm\_simulator provided by IBM Qiskit and evaluated each expectation value by $10^5$ samplings. Each error bar corresponds to statistical errors. The quantum circuits we used are shown in Fig.~\ref{fig:operation}. It should be noticed that measurements of $X$ and $Z$ are done independently since $X$ does not commute with $Z$. The complete data table is given in Table~\ref{tab:processor}, which has energy expectation values of two receivers at $i=1$ and $i=2$ in a corresponding unit lattice. The sender of energy is assigned to $i=0$. The measurements of each operator at different lattice sites $i=1,2$ were done simultaneously. From the table, it is clear that the measurements at $i=1$ do not affect to the results at $i=2$ and vice versa, as predicted. Therefore all receivers in a unit lattice can receive the same amount of energy statistically. In this way, we can realize the homogeneous distribution of energy, without constraints on the spatial distance between a receiver and a sender. 

\begin{table*}
    \centering
    \begin{tabular}{|c|c|c|c|c|c|c|c|}\toprule
      \multicolumn{2}{|c|}{$(h,k)$} &(9,2) &(8,2) & (7,2) & (6,2)  \\\hline
       \multirow{3}{*}{$\langle E_{0}\rangle$}&\{3,6\}& $7.8897\pm0.0090$ & $6.7546\pm0.0080$& $5.5674\pm0.0070$ &$4.3751\pm0.0060$ \\
       &\{3,7\}& $7.6482\pm0.0090$& $6.4994\pm0.0080$ & $5.2865\pm0.0070$ &$4.0341\pm0.0060$ \\
       &\{3,10\}& $6.9239\pm0.0090$ & $5.6837\pm0.0080$ & $4.4021\pm0.0070$ &$3.0898\pm0.0060$ \\\hline
       \multirow{3}{*}{$\langle H_{X,1}\rangle$}&\{3,6\}& $-0.6715\pm0.0037$ & $-0.6956\pm0.0036$ & $-0.7141\pm0.0035$ &$-0.6966\pm0.0033$ \\
       &\{3,7\}& $-0.6497\pm0.0036$ & $-0.6695\pm0.0036$ & $-0.6679\pm0.0035$ &$-0.6357\pm0.0033$ \\
       &\{3,10\}&$-0.5821\pm0.0037$ & $-0.5747\pm0.0036$ & $-0.5372\pm0.0035$ &$-0.4520\pm0.0033$ \\
       \hline
       \multirow{3}{*}{$\langle H_{Z,1}\rangle$}&\{3,6\}&$0.5102\pm0.0037$ & $0.5441\pm0.0036$ & $0.5585\pm0.0035$ &$0.5582\pm0.0034$ \\
       &\{3,7\}& $0.5065\pm0.0036$& $0.5289\pm0.0037$ & $0.5370\pm0.0036$ &$0.5136\pm0.0037$ \\
       &\{3,10\}& $0.4669\pm0.0036$ &$0.4701\pm0.0036$ & $0.4447\pm0.0035$ &$0.3880\pm0.0034$ \\\hline
       \multirow{3}{*}{$\langle E_1\rangle$}&$\{3,6\}$& $-0.1613\pm0.0052$ & $-0.1514\pm0.0051$ & $-0.1556\pm0.0049$ &$-0.1383\pm0.0047$ \\
       &\{3,7\}& $-0.1432\pm0.0051$ & $-0.1406\pm0.0052$ & $-0.1309\pm0.0050$ &$-0.1221\pm0.0049$ \\
       &\{3,10\}& $-0.1151\pm0.0052$ & $-0.1046\pm0.0051$ & $-0.0925\pm0.0049$ &$-0.0640\pm0.0047$ \\
       \hline
       \multirow{3}{*}{$\langle H_{X,2}\rangle$}&\{3,6\}& $-0.6747\pm0.0037$ & $-0.7034\pm0.0036$ & $-0.7073\pm0.0035$ &$-0.6996\pm0.0033$ \\
       &\{3,7\}& $-0.6497\pm0.0037$& $-0.6685\pm0.0036$ & $-0.6674\pm0.0035$ &$-0.6358\pm0.0033$ \\
       &\{3,10\}& $-0.5797\pm0.0037$ & $-0.5752\pm0.0036$ & $-0.5387\pm0.0035$ &$-0.4549\pm0.0033$ \\
       \hline
       \multirow{3}{*}{$\langle H_{Z,2}\rangle$}&\{3,6\}& $0.5169\pm0.0037$ & $0.5467\pm0.0036$& $0.5643\pm0.0035$ &$0.5625\pm0.0034$ \\
       &\{3,7\}& $0.5097\pm0.0036$& $0.5305\pm0.0037$ & $0.5347\pm0.0036$ &$0.5159\pm0.0037$ \\
       &\{3,10\}& $0.4591\pm0.0037$ & $0.4695\pm0.0036$ & $0.4484\pm0.0035$ &$0.3927\pm0.0034$ \\\hline
       \multirow{3}{*}{$\langle E_2\rangle$}&\{3,6\}& $-0.1578\pm0.0052$ & $-0.1567\pm0.0051$ & $-0.1430\pm0.0049$ &$-0.1371\pm0.0047$ \\
       &\{3,7\}& $-0.1400\pm0.0052$ & $-0.1380\pm0.0052$ & $-0.1327\pm0.0050$ &$-0.1199\pm0.0049$ \\
       &\{3,10\}& $-0.1205\pm0.0052$ & $-0.1057\pm0.0051$ & $-0.0903\pm0.0049$ &$-0.0622\pm0.0047$ \\
       \hline
    \end{tabular}
    \caption{Measurement results for each operator. For each measurement, $10^6$ were sampled with a simulator provided by IBM Qiskit}
    \label{tab:processor}
\end{table*}

\section{Implications for our real world}
The hybrid protocol of QST and QET/QED allows energy transfer to remote locations without spatial distance limitations. The application of hyperbolic geometry to information processing systems is quite natural, just as modern classical cyberspace has a hyperbolic geometry. Motivated by this, we have confirmed by simulation that the QED/QET protocol can indeed distribute energy uniformly and simultaneously to multiple nodes by implementing it on a hyperbolic graph. By using QST to relay multiple nodes, energy sent from any point on the graph can be received regardless of distance. In the present day, when hyperbolic geometric quantum processors~\cite{kollar2019hyperbolic} are already established and QST has reached a nearly practical level on various quantum networks~\cite{2021arXiv210112742D,kimble2008quantum,chen2021integrated,doi:10.1126/science.abg1919}, we are ready to verify our long-range QET/QED protocols in a real quantum network. We hope that the results of this study will motivate many experimentalists to undertake demonstrations of quantum energy teleportation and that they will deepen the discussion on how to actually extract the energy and, if so, how it can be used. This will increase the potential for further development of quantum technology and communication. In this regard, it is an important step that the prediction of quantum energy teleportation itself has already been verified in real quantum devices~\cite{2023arXiv230102666I,2022arXiv220316269R}. 

If large-scale quantum energy teleportation is realized, the role of distributing energy at the relay points of a hyperbolic graph foreshadows the emergence of a new business in the quantum era~
\cite{2020QuIP...19...25I,ikeda2022quantum,2021QuIP...20..387I,2022arXiv221102073I,2021QuIP...20..313I}.

\section*{Acknowledgement}
I thank Adrien Florio, David Frenklakh, Sebastian Grieninger, Fangcheng He, Masahiro Hotta, Dmitri Kharzeev, Yuta Kikuchi, Vladimir Korepin, Qiang Li, Adam Lowe, Ren\'{e} Meyer, Shuzhe Shi, Hiroki Sukeno, Tzu-Chieh Wei, Kwangmin Yu and Ismail Zahed for fruitful communication and collaboration. I thank Megumi Ikeda for providing the cartoons. I acknowledge the use of IBM quantum simulators. I was supported by the U.S. Department of Energy, Office of Science, National Quantum Information Science Research Centers, Co-design Center for Quantum Advantage (C2QA) under Contract No.DESC0012704.

\section*{Author contribution}
All work was performed by the author.

\section*{Competing interests}
The author declares that there is no competing financial interests. 

\section*{References}
\bibliographystyle{apsrev4-1.bst}
\bibliography{ref}
\end{document}